\documentclass{appolb}
\usepackage[T1]{fontenc}
\usepackage{lmodern}
\usepackage{amsmath}
\usepackage{graphicx}
\usepackage{booktabs}
\usepackage{xcolor}
\usepackage{cite}
\usepackage[section,above]{placeins}
\usepackage[colorlinks=true,linkcolor=blue!60!black,citecolor=blue!60!black,urlcolor=blue!60!black]{hyperref}
\usepackage{orcidlink}

\newcommand{\mup}{\ensuremath{\mu^+}}
\newcommand{\pip}{\ensuremath{\pi^+}}
\newcommand{\INCL}{\textsc{qgsp\_inclxx}}
\begin{document}
\title{Surface-muon production for ultraslow-muon sources\\ of \mup-based colliders%
\thanks{Contribution to the proceedings of the Workshop on Ultra-cold muons for the LHC, AGH University of Krak\'ow, September 24--26, 2026.}%
}
\author{Kubilay Ince\,\orcidlink{0009-0002-1808-0646}\thanks{Corresponding author: kubilayince@ankara.edu.tr}, Umit Kaya\,\orcidlink{0000-0003-0823-3848}
\address{Institute of Accelerator Technologies, Ankara University, Ankara, T\"urkiye}
}
\headtitle{Surface-muon production for ultraslow-muon sources \ldots}
\headauthor{K.~Ince, U.~Kaya}
\maketitle

\begin{abstract}
Anti-muon colliders based on ultraslow muons, such as $\mu$TRISTAN and the $\mu^+p$ collider LH$\mu$C, need muon intensities well beyond those of present facilities. We study surface-\mup\ production by 3~GeV protons in graphite, molybdenum and tungsten targets with Geant4, for single, tilted and multi-layer target geometries. As single targets, molybdenum and tungsten give comparable yields, more than twice that of graphite, because a larger fraction of the beam interacts. Tilting a target with respect to a narrow beam and splitting it into several layers both raise the yield substantially: five-layer Mo and W targets give 6--8 times the surface-muon yield quoted for the J-PARC MLF graphite target, counted at the production target before capture and transport.
\end{abstract}

\section{Introduction}
Anti-muon colliders avoid ionisation cooling by starting from a low-emittance beam of ultraslow muons (USM): $\mu$TRISTAN~\cite{Hamada2022}, and $\mu^+e^-$ and $\mu^+p$ colliders that combine a muon ring with FCC-ee or CEPC, the HL-LHC ($\mu$LHC), etc.~\cite{Akturk2024EPL,Akturk2025EPJP,Akturk2026muLHC}. In the LH$\mu$C concept of the HELP4CERN Study Group~\cite{Piotrzkowski2026}, 1 or 2~TeV anti-muons collide with the 7~TeV protons of the HL-LHC at centre-of-mass energies of 5.3 or 7.5~TeV and an average luminosity close to $10^{34}$~cm$^{-2}$\,s$^{-1}$, reusing the energy-recovery linac infrastructure proposed for the LHeC~\cite{LHeC2012,LHeC2021}. Its muon source follows $\mu$TRISTAN~\cite{Hamada2022}: a 500~MeV proton linac and a 3~GeV rapid-cycling synchrotron drive a pion-production ring with a cascaded tungsten target that stops the pions. In the USM technique~\cite{Nagamine1995,Kamioka2026review}, surface \mup\ are stopped in silica aerogel, thermal muonium emitted into vacuum is laser-ionised, and the obtained muons are re-accelerated~\cite{Beer2014,Beare2020,Zhang2022}; RF acceleration of such muons was demonstrated at J-PARC in 2024~\cite{Aritome2025}.

About 1\% of the incident surface muons emerge as thermal muonium~\cite{Zhang2022}, and up to about 10\% of these can be ionised with the planned laser system~\cite{Kamioka2026JPS}. For a given beamline and emitter, the USM intensity is therefore proportional to the surface-\mup\ intensity from the production target, provided that the extracted surface muons have a similar phase space; multi-layer muonium emitters raise the conversion efficiency but do not change this proportionality~\cite{Zhang2022,Sakaki2026}. More than $10^{10}\,\mu^+$/s are needed for colliders, which calls for a dedicated production target~\cite{Hamada2022,Sakaki2026,Kamioka2026review}. For the J-PARC MLF target, 20~mm of isotropic graphite~\cite{Miyake2010,Makimura2015}, the $\mu$TRISTAN proposal~\cite{Hamada2022} quotes $1.5\times10^{-3}$ stopped \pip\ and $3\times10^{-5}$ surface \mup\ per 3~GeV proton, of which $1\times10^{-7}$ reach the experiment after the 108~mSr capture solid angle and a transport efficiency of 1/3. That target is kept thin because it must transmit most of the beam to the downstream spallation neutron source. A dedicated target is free of this constraint, so material and geometry can be chosen for the surface-muon yield, as studied at PSI for 590~MeV protons~\cite{Berg2016}.

Sakaki et al.~\cite{Sakaki2026} recently evaluated the hybrid-target concept of $\mu$TRISTAN~\cite{Hamada2022} with a liquid-lithium pion-production target in a recirculating proton ring, surrounded by a metre-scale stack of 0.1~mm tungsten layers that stop the pions and act directly as hot-tungsten muonium emitters. Using PHITS with the JAM model corrected to HARP pion-production data~\cite{HARP2008}, they estimate about $10^{-3}$ ultraslow muons per p--Li collision before extraction, for unit muon-to-muonium conversion efficiency.

In this paper, we use Geant4~\cite{Agostinelli2003,Allison2006,Allison2016} to study the surface-muon yield of compact solid production targets, the conventional alternative in which the muons are transported to the emitter by a surface-muon beamline; the beamline and the emitter are not simulated, so all yields refer to the production target, before capture and transport. We simulate 3~GeV protons on graphite, molybdenum and tungsten in single, tilted and five-layer configurations.

\section{Simulation}
\begin{figure}[!htb]
\centering
\includegraphics[width=\textwidth]{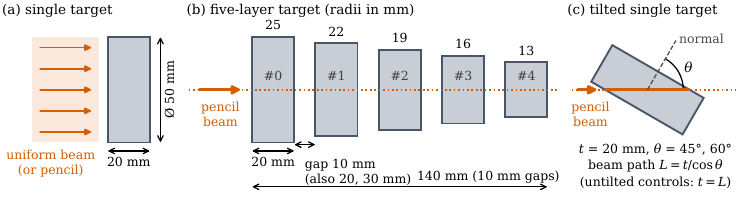}
\caption{Target configurations: (a) single target, (b) five-layer target, (c) tilted single target (top view).}
\label{fig:geom}
\end{figure}

We use Geant4 11.4.2 with the reference physics list \INCL\ and a 0.05~mm production cut. Since no energy here exceeds 3~GeV, the inelastic interactions of the 3~GeV primary protons and of the secondary nucleons and pions are all described by the Li\`ege intranuclear cascade INCL++~\cite{Boudard2013}. All protons have 3~GeV kinetic energy and start in vacuum in front of the target, moving along $+z$ without angular divergence. Two beam profiles are used: a \emph{uniform beam}, with start positions distributed uniformly over a disc of 24.9~mm radius, which covers the 25~mm target face, and a \emph{pencil beam} of zero size on the axis. The targets are shown in Fig.~\ref{fig:geom}: a single cylinder of 25~mm radius and 20~mm thickness of graphite (1.8~g/cm$^3$), molybdenum (10.22~g/cm$^3$) or tungsten (19.3~g/cm$^3$) (Fig.~\ref{fig:geom}(a)), and a five-layer target of coaxial 20~mm layers of the same materials with radii 25, 22, 19, 16 and 13~mm and gaps of 10, 20 or 30~mm (Fig.~\ref{fig:geom}(b)). The single target is irradiated by either beam. The five-layer target is irradiated only by the pencil beam: with a beam as wide as the first layer, much of the beam would hit the layers near their edges or miss the smaller downstream layers, and the protons and secondaries scattered there would leave the stack. To study tilting, the single target hit by the pencil beam is also rotated by $\theta=45^\circ$ and $60^\circ$ about a vertical axis through its centre (Fig.~\ref{fig:geom}(c)). This lengthens the beam path in the target from $t=20$~mm to $L=t/\cos\theta=28.3$ and 40~mm; untilted single targets of these thicknesses serve as controls.

A \emph{surface \mup} is a \mup\ from \pip\ decay at rest that leaves the target; for the five-layer targets it must also escape the layer system without stopping in another layer. Muons from pion decay in flight are excluded; they are 2--9\% of the \mup\ below 30~MeV leaving a single target. To reduce the computing time, e$^\pm$, $\gamma$, neutrinos, $\mu^-$, nuclear fragments, and p, n and $\pi^-$ below 100~MeV are not tracked, and \mup\ born deeper than 1.5~mm are discarded. All configurations are simulated with the same code and Geant4 version. Each configuration is simulated with $5\times10^6$ to $10^8$ protons, the largest numbers for graphite, which has the lowest yield, so that the statistical uncertainty of the surface-\mup\ yield is 1--4\%; only statistical uncertainties are quoted. In every run the number of decay-at-rest \mup\ equals the number of stopped \pip.

\section{Results}
\begin{figure}[!htb]
\centering
\includegraphics[width=\textwidth]{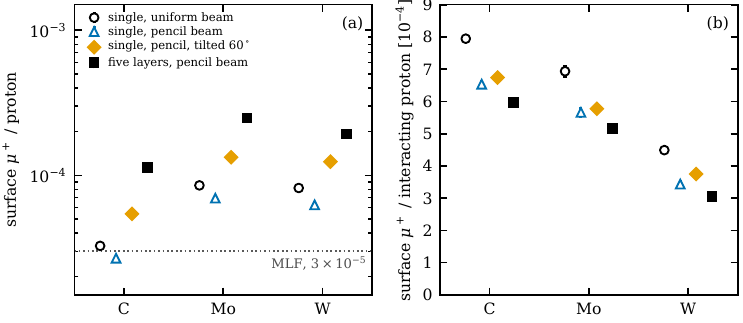}
\caption{Surface-\mup\ yield (a) per proton and (b) per interacting proton. Dotted line: J-PARC MLF value~\cite{Hamada2022}.}
\label{fig:yields}
\end{figure}

The yields are given in Table~\ref{tab:yields} and Fig.~\ref{fig:yields}(a). As single targets hit by the uniform beam, Mo and W give 2.6 and 2.5 times the surface-\mup\ yield of graphite. Tungsten stops 1.6--1.8 times more \pip\ than molybdenum, but its shorter muon range offsets this: W gives 0.96 times the Mo yield as a single target and 0.78 times as a five-layer target.

The advantage of heavy targets comes from using more of the beam. With nuclear interaction lengths of 47.7, 15.2 and 9.9~cm for C, Mo and W~\cite{PDG2024}, 20~mm targets interact 4\%, 12\% and 18\% of the protons, and five-layer targets 19\%, 48\% and 63\% ($P_\mathrm{int}$ in Table~\ref{tab:yields}). Per interacting proton, the surface-\mup\ yield of Mo and W is 0.5--0.9 times that of graphite (Fig.~\ref{fig:yields}(b)), so their higher yield per incident proton comes from the larger number of interactions in a compact target, at the cost of more deposited power. These $(3.0\text{--}7.9)\times10^{-4}$ surface \mup\ per interaction precede beamline transport and muonium conversion, unlike the $\sim10^{-3}$ ultraslow muons per collision of Ref.~\cite{Sakaki2026}, which assume unit conversion efficiency in hot tungsten and include elastic collisions.

Relative to the $3\times10^{-5}$ surface \mup\ per proton quoted for the MLF target~\cite{Hamada2022} (Fig.~\ref{fig:yields}(a)), single Mo and W targets give 2.7--2.8 times more and five-layer Mo and W targets 6.4--8.3 times more; our graphite target gives 1.09 times the reference and 78\% of the quoted $1.5\times10^{-3}$ stopped \pip.

\begin{table}[!htb]
\centering
\caption{Stopped-\pip\ and surface-\mup\ yields. $P_\mathrm{int}$: fraction of protons interacting along the beam path~\cite{PDG2024}.}
\label{tab:yields}
\footnotesize\setlength{\tabcolsep}{4pt}
\begin{tabular}{llcccc}
\toprule
 & & & stopped \pip/p & surface \mup/p & \mup/int.\\
Config. & Mat. & $P_\mathrm{int}$ & ($10^{-3}$) & ($10^{-5}$) & ($10^{-4}$)\\
\midrule
single layer,      & C  & 0.041 & 1.17  & $3.27\pm0.06$ & 7.9\\
uniform beam       & Mo & 0.123 & 13.3  & $8.53\pm0.24$ & 6.9\\
                   & W  & 0.182 & 23.8  & $8.18\pm0.29$ & 4.5\\
single layer,      & C  & 0.041 & 1.38  & $2.69\pm0.05$ & 6.5\\
pencil beam        & Mo & 0.123 & 15.9  & $6.96\pm0.22$ & 5.7\\
                   & W  & 0.182 & 27.9  & $6.26\pm0.25$ & 3.4\\
single layer,      & C  & 0.080 & 2.71  & $5.43\pm0.07$ & 6.7\\
pencil beam,       & Mo & 0.231 & 29.9  & $13.3\pm0.3$  & 5.8\\
tilted $60^\circ$  & W  & 0.331 & 51.6  & $12.4\pm0.4$  & 3.7\\
five layers,       & C  & 0.189 & 7.28  & $11.3\pm0.1$  & 6.0\\
pencil beam,       & Mo & 0.481 & 71.1  & $24.8\pm0.7$  & 5.2\\
10~mm gaps         & W  & 0.634 & 112   & $19.3\pm0.6$  & 3.0\\
\bottomrule
\end{tabular}
\end{table}

\begin{figure}[!htb]
\centering
\includegraphics[width=\textwidth]{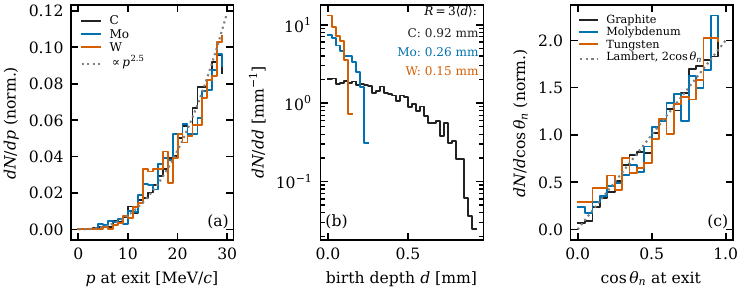}
\caption{Single targets, uniform beam: (a) momentum spectra at exit, (b) birth depth, (c) exit direction. Dotted: $p^{2.5}$ and Lambert distribution.}
\label{fig:escape}
\end{figure}

The birth depth $d$ of the escaping muons (Fig.~\ref{fig:escape}(b)) falls roughly linearly to zero at the muon range $R$, as expected for isotropic emission from uniformly distributed stopped pions, for which $\langle d\rangle=R/3$. This gives $R=0.92$~mm (C), 0.26~mm (Mo) and 0.15~mm (W), consistent with the range derived from the ratio of escaping muons to pions stopped within 0.5~mm of the surface (Table~\ref{tab:escape}). The exit directions follow a Lambert ($\cos\theta$) distribution about the surface normal (Fig.~\ref{fig:escape}(c)). The yield is thus set by the stopped-\pip\ density within one range of the surface, and the longer range in Mo (0.26~mm against 0.15~mm) offsets its lower pion density relative to W. The momentum spectra (Fig.~\ref{fig:escape}(a)) follow $dN/dp\propto p^{n}$ with $n=2.2$--2.3 for single targets (fit range 8--29.8~MeV/$c$, the upper limit being the muon momentum from \pip\ decay at rest; Table~\ref{tab:escape}), close to $n=2.5$ expected from $R\propto p^{3.5}$~\cite{Pifer1976}. Over all 27 simulated configurations, including the tilted and thicker targets discussed below, $n=2.0$--2.6, $\langle p\rangle\approx23$~MeV/$c$, and 42--50\% of the muons are above 25~MeV/$c$.

\begin{table}[!htb]
\centering
\caption{Surface-\mup\ properties: mean exit momentum, spectral index $n$ ($dN/dp\propto p^{n}$, 8--29.8~MeV/$c$), fraction $f$ above 25~MeV/$c$, muon range $R=3\langle d\rangle$ and $R_\mathrm{ratio}$, and fraction from pion decay in flight. Single targets: uniform beam.}
\label{tab:escape}
\footnotesize\setlength{\tabcolsep}{4pt}
\begin{tabular}{lcccccc}
\toprule
Target & $\langle p\rangle$ (MeV/$c$) & $n$ & $f$ & $R$ (mm) & $R_\mathrm{ratio}$ (mm) & in flight\\
\midrule
C  & 22.9 & $2.32\pm0.07$ & 0.43 & 0.92 & 0.92 & 0.090\\
Mo & 22.6 & $2.16\pm0.10$ & 0.43 & 0.26 & 0.23 & 0.048\\
W  & 22.9 & $2.18\pm0.13$ & 0.45 & 0.15 & 0.13 & 0.022\\
\midrule
five layers & 23.2 & 2.48--2.64 & 0.45--0.48 & & & \\
\bottomrule
\end{tabular}
\end{table}

\begin{figure}[!htb]
\centering
\includegraphics[width=\textwidth]{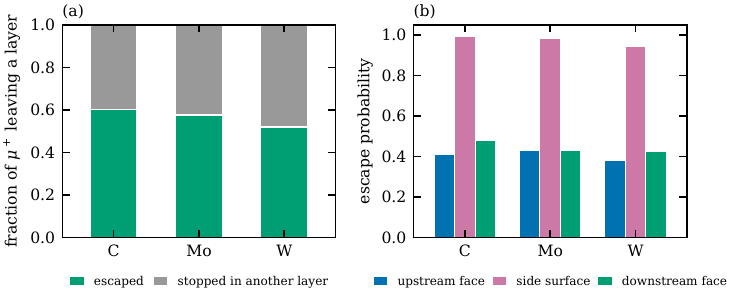}
\caption{Five-layer targets: (a) fate of surface \mup\ leaving a layer, (b) escape probability by exit face.}
\label{fig:multi}
\end{figure}

\begin{figure}[!htb]
\centering
\includegraphics[width=\textwidth]{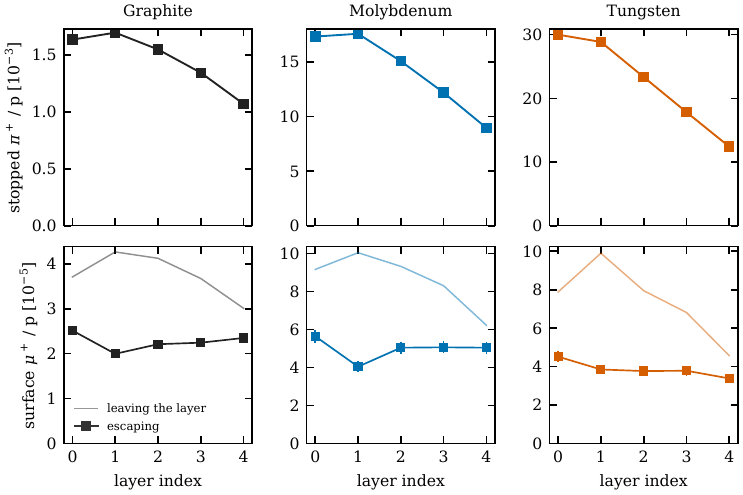}
\caption{Five-layer targets, per layer: stopped \pip\ (top) and surface \mup\ (bottom).}
\label{fig:layers}
\end{figure}

\begin{table}[!htb]
\centering
\caption{Five-layer targets with 20 and 30~mm gaps, relative to 10~mm gaps, and escape fraction of the \mup\ leaving a layer.}
\label{tab:gap}
\footnotesize\setlength{\tabcolsep}{4pt}
\begin{tabular}{lccccccc}
\toprule
 & \multicolumn{2}{c}{surface \mup} & \multicolumn{2}{c}{stopped \pip} & \multicolumn{3}{c}{escape fraction}\\
\cmidrule(lr){2-3}\cmidrule(lr){4-5}\cmidrule(lr){6-8}
Mat. & 20~mm & 30~mm & 20~mm & 30~mm & 10~mm & 20~mm & 30~mm\\
\midrule
C  & $1.06\pm0.01$ & $1.08\pm0.01$ & 0.92 & 0.87 & 0.60 & 0.74 & 0.84\\
Mo & $1.06\pm0.04$ & $1.07\pm0.04$ & 0.93 & 0.89 & 0.58 & 0.72 & 0.82\\
W  & $1.16\pm0.05$ & $1.23\pm0.05$ & 0.93 & 0.89 & 0.52 & 0.68 & 0.79\\
\bottomrule
\end{tabular}
\end{table}

Splitting the target into five layers with 10~mm gaps multiplies the emitting surface. The escaping yield per proton is $(11.3\pm0.1)\times10^{-5}$ for graphite, $(24.8\pm0.7)\times10^{-5}$ for Mo and $(19.3\pm0.6)\times10^{-5}$ for W (Table~\ref{tab:yields}), i.e.\ 3.1--4.2 times the single target of the same material hit by the same pencil beam (C 4.2, Mo 3.6, W 3.1). On the single target, the pencil beam stops 17--19\% more \pip\ than the uniform beam, because they are produced closer to the axis, but gives 18--23\% fewer surface \mup, because fewer of them stop near the side surface; relative to the uniform-beam single target the five-layer gain is therefore 2.4--3.5. Concerning the muons leaving a layer, 39--48\% of them stop in a neighbouring layer (Fig.~\ref{fig:multi}(a)): only 38--48\% of those leaving through a face escape, compared with 94--99\% of those leaving through a side surface (Fig.~\ref{fig:multi}(b)). Accordingly, the escape fraction is lowest for the second layer (0.39--0.47) and highest for the last (0.74--0.81), whose downstream face is unobstructed (Fig.~\ref{fig:layers}). The stopped-\pip\ yield falls along the stack: in the last layer it is 0.65 of the first-layer value for graphite and 0.41--0.52 for Mo and W, where only 59\% and 45\% of the protons reach the last layer without interacting. The rising escape fraction partly compensates, and the first layer gives the largest escaping yield in all targets (W: $4.5\times10^{-5}$ escaping \mup\ per proton in the first layer, $3.4\times10^{-5}$ in the last).

Widening the gaps from 10 to 20 and 30~mm raises the escape probability of the muons leaving a layer from 0.52--0.60 to 0.68--0.74 and 0.79--0.84, but more pions now leave the stack through the gaps, so that 7--8\% and 11--13\% fewer \pip\ stop. The escaping yield therefore increases by only 6--16\% and 7--23\% (Table~\ref{tab:gap}), for a stack 180 and 220~mm long instead of 140~mm.

\begin{table}[!htb]
\centering
\caption{Tilted and thicker single targets (pencil beam), relative to the untilted 20~mm target.}
\label{tab:tilt}
\footnotesize\setlength{\tabcolsep}{3.6pt}
\begin{tabular}{lcccccccc}
\toprule
 & \multicolumn{4}{c}{surface \mup} & \multicolumn{4}{c}{stopped \pip}\\
\cmidrule(lr){2-5}\cmidrule(lr){6-9}
 & \multicolumn{2}{c}{tilted, $\theta$} & \multicolumn{2}{c}{untilted, $t$ (mm)} & \multicolumn{2}{c}{tilted, $\theta$} & \multicolumn{2}{c}{untilted, $t$ (mm)}\\
\cmidrule(lr){2-3}\cmidrule(lr){4-5}\cmidrule(lr){6-7}\cmidrule(lr){8-9}
Mat. & $45^\circ$ & $60^\circ$ & 28.3 & 40 & $45^\circ$ & $60^\circ$ & 28.3 & 40\\
\midrule
C  & $1.46\pm0.04$ & $2.02\pm0.05$ & $1.24\pm0.03$ & $1.56\pm0.04$ & 1.42 & 1.97 & 1.62 & 2.56\\
Mo & $1.45\pm0.06$ & $1.91\pm0.07$ & $1.17\pm0.05$ & $1.32\pm0.05$ & 1.39 & 1.88 & 1.56 & 2.35\\
W  & $1.36\pm0.07$ & $1.98\pm0.10$ & $1.14\pm0.06$ & $1.30\pm0.07$ & 1.38 & 1.85 & 1.52 & 2.23\\
\bottomrule
\end{tabular}
\end{table}

A single target can also be tilted. For the pencil beam, tilting the 20~mm target by $\theta=45^\circ$ and $60^\circ$ raises the surface-\mup\ yield by factors of 1.36--1.46 and 1.91--2.02 (Table~\ref{tab:tilt}), close to $1/\cos\theta=1.41$ and 2.00, for all three materials. The stopped-\pip\ yield rises by about the same factors (1.38--1.42 and 1.85--1.97), so the fraction of stopped pions that give a surface muon hardly changes: the beam path per unit depth, and with it the density of stopped pions near the two faces, grows by $1/\cos\theta$. Untilted targets with the same beam path stop more pions, 1.52--1.62 and 2.23--2.56 times the 20~mm value, but give only 1.14--1.24 and 1.30--1.56 times as many surface \mup, because the additional pions stop far from the faces. The tilted target therefore gives 1.2 ($45^\circ$) and 1.3--1.5 ($60^\circ$) times the yield of the untilted target with the same beam path; per interacting proton, it is at least as efficient as the untilted 20~mm target (Fig.~\ref{fig:yields}(b)). At $60^\circ$, Mo and W give $(13.3\pm0.3)\times10^{-5}$ and $(12.4\pm0.4)\times10^{-5}$ surface \mup\ per proton (Table~\ref{tab:yields}), 4.1--4.4 times the MLF reference value, and all three materials give 1.5--1.7 times the yield of the uniform-beam single target. The gain requires a beam that is narrow compared with the target face. For a uniform beam wider than the target, tilting lengthens the path of each proton by $1/\cos\theta$ but reduces the number of protons hitting the target by about $\cos\theta$, so that the total beam path in the target does not change. Berg et al.~\cite{Berg2016} found a gain of 30--60\% for a rotated slab target hit by the narrow ($\sigma\approx1$~mm) proton beam at PSI.

The escape properties found here also apply to schemes that collect the pions produced in a thin target and stop them in a separate stack of foils~\cite{Hamada2022,Sakaki2026}, whereas a surface-muon target uses only the pions that stop within one muon range of a surface, 0.2--3\% of the stopped \pip\ here. For pions stopping uniformly in a foil of thickness $t\geq R$, a fraction $R/(2t)$ of the decay muons leaves the foil, about 7\% for 1~mm tungsten foils, whereas foils thinner than $R\approx0.15$~mm release most of them; the escaping muons have the $p^{2.5}$ spectrum and the Lambert angular distribution obtained above.

\section{Conclusions}
We study surface-\mup\ production by 3~GeV protons in graphite, molybdenum and tungsten targets with the Geant4 physics list \INCL. As single 20~mm targets, Mo and W give 2.5--2.6 times the surface-\mup\ yield of graphite and agree with each other within 4\%: tungsten stops more pions, but the muon range is shorter (0.15~mm against 0.26~mm in Mo and 0.92~mm in graphite). Their advantage over graphite comes from the larger fraction of the beam that interacts in a compact target. A five-layer target gives 3.1--4.2 times the single target hit by the same pencil beam, i.e.\ $(24.8\pm0.7)\times10^{-5}$ and $(19.3\pm0.6)\times10^{-5}$ surface \mup\ per proton for Mo and W, 6--8 times the surface-muon yield quoted for the J-PARC MLF graphite target; 39--48\% of the muons leaving a layer stop in a neighbouring layer, and widening the gaps to 30~mm gains only 7--23\%. Tilting a single target hit by a pencil beam by $45^\circ$ and $60^\circ$ raises the yield by factors of 1.36--1.46 and 1.91--2.02, close to $1/\cos\theta$, for all materials, 1.2 and 1.3--1.5 times more than untilted targets of the same beam path length, because the stopped-pion density near the faces grows in proportion to the beam path; at $60^\circ$, a single 20~mm Mo or W target gives $(13.3\pm0.3)\times10^{-5}$ and $(12.4\pm0.4)\times10^{-5}$ surface \mup\ per proton, respectively. All yields refer to the production target, before capture and transport; the transport efficiency, thermal load and radiation damage are not considered here, and the capture efficiency, which depends on the source geometry, has to be evaluated for the multi-layer and tilted targets. These aspects will be addressed in future work.

\section*{Acknowledgements}
We thank K.~Piotrzkowski and the Krak\'ow group for organising the Workshop on Ultra-cold muons for the LHC and for the opportunity to contribute to its proceedings. We are grateful to M.~Otani and C.~Zhang for useful discussions. We also thank Ali Can Canbay, Saleh Sultansoy and Burak Da\u{g}l\i\ for their support.

\bibliographystyle{unsrt}
{\small
\bibliography{refs}}

\end{document}